\documentclass[11pt,twoside]{article}

\usepackage{asp2006}

\def\rmit#1{{\it #1}}              %% italics (RR mode, Kluwer)
\def\specchar#1{{\sc #1}}
\def\FeI{\mbox{Fe\,\specchar{i}}}

\def\SiI{\mbox{Si\,\specchar{i}}}
\def\HeI{\mbox{He\,\specchar{i}}}
\def\NiI{\mbox{Ni\,\specchar{i}}}

\def\ie{\rmit{i.e.}}
\def\eg{\rmit{e.g.}}

\begin{document}
\title{Simulations of waves in sunspots}   %%% Fill in title
\author{Elena Khomenko}   %%% Fill in author names

\affil{Instituto de Astrof\'{\i}sica de Canarias, 38205, C/
V\'{\i}a L{\'a}ctea, s/n, Tenerife, Spain; \\
 Main Astronomical Observatory, NAS, 03680, Kyiv, Ukraine}    %%% Fill in author affiliations

\begin{abstract}
Magnetic field modifies the properties of waves in a complex way.
Significant advances has been made recently in our understanding
of the physics of waves in solar active regions with the help of
analytical theories, numerical simulations, as well as
hi-resolution observations. In this contribution we review the
current ideas in the field, with the emphasis on theoretical
models of waves in sunspots.
\end{abstract}

%%% MAIN BODY OF TEXT GOES HERE. CONSULT "INSTRUCTIONS FOR AUTHORS USING
%%% LATEX2E MARKUP", SECTIONS 2.3-2.6 FOR HELP WITH EQUATIONS, FIGURES,
%%% AND TABLES.

\section{Motivation and questions}   %%% Top level section head (remove "%" symbol)

Photospheres and chromospheres of sunspots are the regions where
different physical agents come into play with almost equal weight.
The restoring forces in the wave equation, such as magnetic
Lorentz force, gas pressure gradient and buoyancy, are of the same
order of magnitude, making arbitrary the division into pure wave
modes and complicating theoretical models. A recent summary on the
observational properties of sunspot oscillations together with
some theoretical issues is presented by \citet{Bogdan+Judge2006}.
The most recent observational results suggest that wave phenomena
at different layers of the sunspots umbrae and penumbrae are
related with each other \citep[see \eg][]{Marco+etal1996,
Maltby+etal1999, Maltby+etal2001, Brynildsen+etal2000,
Brynildsen+etal2002, Christopoulou+etal2000,
Christopoulou+etal2001, Rouppe+etal2003, Tziotziou+etal2006,
Bloomfield+etal2007}. The interpretation of the observational
material defines several groups of questions concerning the
sunspot wave physics that can be summarized in the following way:
\begin{itemize}
\item What drives the waves observed in sunspots? Are they
externally driven by the quiet Sun $p$-modes? Are there sources of
oscillations inside the umbra due to weak convection?
\item How can the oscillations observed at different sunspot layers
be interpreted in terms of MHD waves? What are the relationships
between photospheric and chromospheric oscillations? What causes
the complex spatial pattern of oscillations such as chromospheric
umbral flashes, penumbral waves, spatial coherency of waves over
the umbra, etc.?
\item Why is the wave power suppressed in the umbral photosphere
compared to the quiet Sun? Why is it enhanced in the chromosphere?
\item What mechanisms produce the change of the dominating frequency of
waves in the umbra from 3 mHz in the photosphere to 5--6 mHz in
the chromosphere? What are the reasons for the spatial power
distribution at different frequency intervals observed in the
umbra and penumbra?
\item Are there observational evidences of the mode transformation
in sunspots?
\item What is the source of helioseismological velocity signal
detected in active regions? Are they due to fast, slow MHD waves?
What are the consequences of the strong magnetic field of sunspots
onto helioseismology measurements and inversions of
sub-photospheric structure of sunspots?
\end{itemize}
Some of these questions, like the interpretation of the wave modes
or change of the frequency with height, received larger attention
in the literature in the past and can be considered answered to a
smaller or larger extent. Some others, like those concerning local
helioseismology in active regions, only received attention
recently and no unique answer has been found yet. In the rest of
paper we will describe the existing models and the theoretical
progress made in understanding sunspot oscillations.

\section{Excitation of oscillations in sunspots}

It is generally accepted that the quiet-Sun oscillations are
stochastically excited by solar convection at the top part of the
convection zone \citep{Goldreich+Kumar1990, Nordlund+Stein2001}.
The frequency spectrum and the temporal behaviour of waves in the
umbra is very similar to the quiet Sun, except for reduced power
\citep[\eg ][]{Penn+LaBonte1993}. It is systematically found in
observations that more wave power travels toward the umbra than
leaves the umbra. One of the natural explanation of this effect is
to assume that the sunspot waves are driven by external $p$-modes,
modified inside the magnetized sunspot atmosphere.

A number of theoretical investigations has been performed assuming
the incoming $p$-mode waves incident on a sunspot-like magnetic
field concentration \citep[\eg\ ][]{Cally+Bogdan+Zweibel1994,
Cally+Bogdan1997, Rosenthal+Julien2000} and studying the wave
transformation, ``absorption'' and scattering by sunspots. These
studies have shown that both amplitudes and phases of the incoming
waves are modified after passing thought the magnetic field
regions. In particular, the surface amplitudes of the $p$-modes
were found to be reduced due to the mode transformation (Sect. 5).

Despite the presence of strong magnetic field in sunspot's umbra
and penumbra, the models of magnetoconvection indicate that these
environments can be convective \citep{Lee1993, Weiss+etal1990,
Cattaneo+Emonet+Weiss2003, Schussler+Vogler2006,
Spruit+Scharmer2006}. The narrow nearly field-free upflowing
umbral dots adjacent by downflows are the result of convection in
a strong magnetic field. Recent observations seem to confirm this
idea \cite[see \eg\ ][]{Bharti+etal2007, Watanabe+etal2008}. Thus,
some waves can be generated inside the sunspot umbra as well.
\citet{Jacoutot+etal2008} performed numerical simulations of
magnetoconvection and studied the spectra of generated waves as a
function of magnetic flux in the model. Apart from the power
suppression in regions with enhanced magnetic field, these
simulations suggest an increase of high-frequency power (above 5
mHz) for intermediate magnetic field strengths (of the order of
300--600 G) caused by changes of the spatial-temporal spectrum of
turbulent convection in a magnetic field.

It can be concluded that both external driving and in-situ
excitation by convection are able explain the observed properties
(\eg\ power reduction) of waves in sunspots. A qualitative
estimation of the effect of the reduced wave excitation in
sunspots and the direct comparison to observations was reported by
\citet{Parchevsky+Kosovichev2007}. The authors performed
hydrodynamical simulations of waves generated by random sources,
with the strength of the sources reduced to zero in the sunspot
umbra. They obtained that, even though no waves were excited in
the umbra, the velocity power measured there was about twice
larger than actually observed. Thus, the waves detected in
sunspots must be a mixture of external $p$-modes (and other wave
types produced by the mode conversion) and MHD waves generated
directly inside the sunspot by the weak convection. The relative
contribution of the different effects is to be determined.

\section{Interpretation of oscillations in terms of MHD waves}

Historically, the sunspot oscillations were divided into 5-min
photospheric oscillations, 3-min chromospheric oscillations and
running penumbral waves \citep{Bogdan+Judge2006}. Recent
observations show that these oscillations can be a manifestation
of the same physical phenomena \citep{Maltby+etal1999,
Maltby+etal2001, Brynildsen+etal2000, Brynildsen+etal2002,
Christopoulou+etal2000, Christopoulou+etal2001, Rouppe+etal2003,
Centeno+etal2006a, Tziotziou+etal2006, Tziotziou+etal2007}.
Chromospheric observations show umbral flashes followed by a
smooth continuous quasi-circular expansion of the visible
perturbation pattern from the umbra through the penumbra toward
the edge of sunspot. At the same time, the frequency of
chromospheric oscillations decreases  from 5 to 3 mHz from the
umbra to the penumbra \citep{Tziotziou+etal2007}. The waves in the
umbra propagate from the photosphere to the chromosphere along the
magnetic field lines forming shocks of about 10 km/sec
peak-to-peak amplitude at heights of formation of \HeI\ 10380 \AA\
line  \citep{Lites1986, LopezAriste+etal2001, Rouppe+etal2003,
Centeno+etal2006a}.

The characteristic speeds of waves change many orders of magnitude
along the photosphere and chromosphere of sunspots. At some height
waves propagate through the region where the sound and the
Alfv\'en speed are equal ($c_S = v_A$). In this region mode
transformation and coupling of different wave phenomena occurs. In
addition, the magnetic field and thermodynamic variables show
important gradients both in horizontal and vertical directions.
All these ingredients make realistic modeling of sunspot waves a
rather difficult task only accessible via numerical simulations.
Theoretical bases of the MHD wave propagation in a stratified
atmosphere with a constant magnetic field were developed by, \eg,
\citet{Ferraro+Plumpton1958, Osterbrock1961,
Zhugzhda+Dzhalilov1982, Zhugzhda+Dzhalilov1984a,
Evans+Roberts1990}. One of the first numerical simulations of
waves in non-trivial magneto-atmospheres with applications to
solar photosphere and chromosphere was performed by
\citet{Rosenthal+etal2002, Bogdan+etal2003}. Similar calculations
for conditions appropriate for sunspots rather than flux tubes
were carried out by \citet{Khomenko+Collados2006}. The waves were
generated by a photospheric pulse located inside the magnetic
field region and propagated through the layer where $c_S = v_A$ on
their way from the photosphere to the chromosphere.
\citet{Rosenthal+etal2002} and \citet{Bogdan+etal2003} pointed out
an important role of the magnetic canopy defined as the region
where the plasma $\beta$ ($\beta=8\pi p/B^2= (\gamma/2)
c_S^2/v_A^2 $) is close to unity. According to their results, the
properties of waves observed in the atmosphere depend sensitively
on the location and orientation of the magnetic canopy. The mode
transformation occurs in the canopy region and several wave types
can be present simultaneously there traveling in different
directions. However, the fast and the slow wave modes decouple
effectively away from the transformation layer. At heights above
$\beta=1$ the refraction and reflection of the fast (magnetic)
mode occur \citep{Rosenthal+etal2002, Khomenko+Collados2006}. The
slow mode (either produced by the driver or after the mode
transformation) is always guided along the magnetic field lines,
except at the region $\beta\approx 1$ where it deviates from this
direction by maximum 27 degrees \citep{Osterbrock1961}.
These and other theoretical investigations demonstrated that the
explanation of the observed oscillations in magnetic structures in
terms of MHD waves requires the knowledge of the location of the
$\beta=1$ level, as well as the inclination of the magnetic field
lines.

\citet{Mathew+etal2004} used spectropolarimetric observations in
\FeI\ 1.56 $\mu$m lines to calculate maps of plasma $\beta$ for
the observed sunspot. The plasma beta at heights of the continuum
formation of radiation at 1.56 $\mu$m was found to be less than
$1$ in most of the umbra and inner penumbra reaching values above
1.5 in the outer penumbra. Most photospheric lines used typically
in observations (like \NiI\ 6768 \AA\ used by SOHO/MDI) are formed
even higher in the atmosphere. It indicates that oscillations
observed in sunspot umbra should be due to the low $\beta$ MHD
waves. The situation is more complex in the penumbra where the
plasma $\beta$ should be around unity at the typical heights of
formation of photospheric spectral lines.

From observations, there is no clear conclusions on what wave
types are detected in sunspots. The analysis of magnetic field
oscillations in a sunspot obtained from spectropolarimetric
observations of \FeI\ 1.56 $\mu$m lines allowed
\citet{Khomenko+Collados+BellotRubio2003} to conclude that the
observed waves are a mixture of fast and slow low-$\beta$ waves
and the contribution of the slow (acoustic-like) wave is the
dominant in the sunspot umbra.
The slow (acoustic-like) wave propagation along the magnetic field
lines from the photosphere to the chromosphere in the sunspot
umbra was reported from spectropolarimetric observations by
\citet{Centeno+etal2006a}.
In the penumbra, \citet{Bloomfield+etal2007a} obtained that the
propagation speeds of waves in the deep photosphere most closely
resemble the fast-mode speeds of the modified $p$-modes, \ie\
high-$\beta$ acoustic-like waves propagating toward the sunspot at
an angle $\sim$ 50 degrees to the vertical. In a subsequent paper,
\citet{Bloomfield+etal2007} used a time series of photospheric
\SiI\ and chromospheric \HeI\ spectra at 1.08 $\mu$m and concluded
that the running penumbral waves are a visible pattern of
low-$\beta$ slow mode waves propagating and expanding their
wavefront along the inclined magnetic field lines in the penumbra.
Similar interpretation of the running penumbral waves was also
suggested by \citet{Bogdan+Judge2006}.
Anticipating the conclusions from Sect. 6 of this paper,
theoretical modeling of local helioseismology signals in sunspot
regions seem to suggest that they are rather due to high-$\beta$
fast mode waves (modified $p$-modes) \citep{Khomenko+etal2008b}.
Given these diverse conclusions, more work is required to unify
the picture of MHD wave modes observed in sunspots.

%Evans and Roberts (1990) consider analytically the oscillations in
%a sunspot-like flux tubes and come to the conclusion that the
%3-min oscillations and slow-body waves driven by convection and
%that the 5-min oscillations observed in the photosphere and below
%are the fast-body waves. The running penumbral waves are fast and
%slow surface waves.
%
%Kshevetskii and Solovev (2008) excite internal gravity waves in
%he atmosphere overlaying sunspots by introducing oscillations of
%the vertical velocity at the lower boundary. These waves propagate
%neatly horizontally and their energy drops exponentially with
%height.

\section{Frequency distribution}

%Mainly observational. Bogdan and Judge explanation of the
%frequency shift from the photosphere to the chromosphere.
%Zhugzhda, and many others - resonance on the temperature
%structure. Fleck and Schmit  - resonance excitation of the
%fundamental mode. Horizontal distribution is unknown.

The power of velocity and intensity oscillations is distributed in
a complex way over the active regions \citep{Hindman+Brown1998,
Jain+Haber2002, Tziotziou+etal2007, Nagashima+etal2007,
Mathew2008}. In the photosphere, the power of oscillations in
generally suppressed at all frequencies, both in the umbra and in
the penumbra, except for the bright ring at the umbra-penumbra
boundary
\citep{Mathew2008}. Acoustic halos are found at the edges
of active regions depending magnetic field strength
\citep{Braun+etal1992, Brown+Bogdan+Lites+Thomas1992,
Hindman+Brown1998} being most prominent at high frequencies above
5 mHz. In the chromosphere, the power is suppressed at low
frequencies (below 2.5 mHz) in the umbra but then is enhanced in
the high frequencies (above 3.5 mHz) \citep[\eg\
][]{Nagashima+etal2007}. The most prominent are 5--6 mHz
oscillations observed in the umbral chromosphere. The running
penumbral waves in the chromosphere have lower frequency around 3
mHz \citep{Tziotziou+etal2007}.

Assuming the waves observed in the umbral photosphere and
chromosphere are slow low-$\beta$ acoustic-like waves propagating
along nearly vertical magnetic field lines, the mechanism of the
frequency variation with height of these waves should be similar
to that of the ordinary acoustic waves in the non-magnetic Sun.
\citet{Fleck+Schmitz1991} have demonstrated that the change of
frequency with height from 3 to 5--6 mHz is due to a resonant
excitation at the atmospheric cut-off frequency, occurring even
for linear waves in an isothermal atmosphere \citep[see
also][]{Kalkofen+etal1994}. In the solar case, the low
temperatures in the upper photosphere give rise to the cut-off
frequency around 5 mHz. The response of the solar atmosphere to
non-linear adiabatic shock wave propagation also leads to an
appearance of the 5 mHz frequency peak in the power spectra, under
the condition the underlying photosphere has a low frequency 3 mHz
component \citep{Fleck+Schmitz1993}.
At the same time, \citet{Zhugzhda+Locans1981,
Gurman+Leibacher1984, Zhugzhda2007} argue that the observed
spectrum of sunspot umbral oscillations in the chromosphere is
rather due to a particular temperature gradients of the atmosphere
acting as an interference filter for linear three-minute period
acoustic waves.
Yet another explanation was suggested by
\citet{Carlsson+Stein1997} based on the fact that at lower
frequencies the energy falls off exponentially with height,
stronger that for the high-frequency waves, not affected by the
cut-off. Similar effect was discussed also by
\citet{Bogdan+Judge2006}.

More puzzling is the horizontal distribution of wave frequencies
over active regions observed at different heights. The power
suppression in the umbral photosphere is thought to be due to the
mode conversion extracting the energy of the incident $p$-mode
waves \citep[\eg][see the discussion in the next
Section]{Cally+Bogdan1997, Cally+Crouch+Braun2003,
Crouch+etal2005}. The power distribution in the chromosphere can
be explained assuming again the low-$\beta$ slow mode propagation
along the inclined magnetic field lines \citep{Bogdan+Judge2006,
Bloomfield+etal2007}. In the low-$\beta$ plasma, the effective
cut-off frequency along each individual magnetic field line
decreases by a cosine of its inclination angle $\theta$. Taking
the typical values of the inclination in the umbra and penumbra,
this mechanism is able to produce the required change from the 5
mHz frequency characteristic for umbral flashes to 2-3 mHz
characteristic for running penumbral waves, as confirmed by
spectropolarimetric observations of \citet{Bloomfield+etal2007}.
As for the high-frequency acoustic halo surrounding active
regions, no plausible explanation exists as of today. In his
recently reported simulations \citet{Hanasoge2008} calculated the
rms wave power after randomly distributing acoustic sources over
the region containing a sunspot-like flux tube. Both power
reduction in the ``umbra'' and halo effect in the surrounding
appeared naturally in these simulations, confirming that they are
definitely an MHD effect. Certainly, the mode conversion and wave
refraction in the inclined magnetic field at the edges of active
regions play a role in the appearance of acoustic halo, requiring
a further study.

\section{Mode transformation}

%Generally speaking the wave transformation produces both slow
%waves extracting energy upwards as well as downwards from the p
%modes.
%Schunker+Cally2006 talk about the importance of the height if
%formation of NiI line, acoustic cut-off height and the
%transformation layer. Argue that in observations of Schunker(2005,
%2006) there is a dependence of the  phase shift of waves on the
%penumbra viewing angle, i.e. the inclination angle of the magnetic
%field lines. Thus, the waves have a preferred direction of
%propagation, and must be slow low-beta waves.%
%Concepts, then consequences (power reduction, acoustic halos),
%then dispersion (larger transformation for certain angles) 3D

Mode transformation causes important effects on the observed wave
propagation in active regions. The bases of the MHD wave
transformation theory in a stratified solar atmosphere were
developed by \citet{Zhugzhda+Dzhalilov1982}. Approaching the layer
where the acoustic and the Alfv\'en speeds are equal, the phase
speeds of the different MHD modes become close and the energy can
be transferred between the different branches of the dispersion
relation. The direction and the effectiveness of the mode
transformation depends, among the other parameters, on the wave
frequency and the attacking angle between the wave vector
$\vec{k}$ and the magnetic field \citep[\eg][]{Cally2005,
Cally2006, Cally+Goossens2008}. In the two-dimensional case, an
approximate formula for the transformation coefficient from  fast
to slow mode was derived by \citet{Cally2005} for the
high-frequency waves (above the acoustic cut-off) in a vertical
magnetic field:
\begin{equation}
T=\exp\left( -k\pi\sin^2\psi /
\left|\frac{d}{dz}\frac{c_S^2}{v_A^2}\right|\right) \,,
\label{eq:cally}
\end{equation}
where $\psi$ is the angle between $\vec{k}$ and the magnetic field
$\vec{B}$. According to this equation, the fast-to-slow mode
transformation is complete ($T=1$) for the waves with $\vec{k}$
directed along $\vec{B}$. For larger angles $\psi$ the efficiency
of the fast-to-slow mode transformation rapidly decreases. The
higher the frequency of waves (implying generally smaller $k$),
the smaller is the cone of $\psi$'s where the fast-to-slow mode
transformation is effective.

Numerical simulations of the $p$ and $f$ mode interaction with a
vertical magnetic field concentration have demonstrated that,
indeed, a significant power  of the incident modes is converted
into the slow high-$\beta$ magneto-acoustic modes propagating
downwards along the magnetic field lines and leading to a visible
power reduction at the surface \citep{Cally+Bogdan+Zweibel1994,
Cally+Bogdan1997, Rosenthal+Julien2000}. In addition, these
simulations have shown that the fraction of the $f$-mode power
transformed to slow modes is sufficient to explain its power
reduction is observations. On the contrary, the $p$ modes were not
transformed sufficiently in the vertical magnetic field in order
to be explained by this mechanism.

Theoretical models of the fast-to-slow mode transformation in the
inclined magnetic field have demonstrated that it can be
particularly strong for a narrow  range of the magnetic field
inclinations around 20--30 degrees to the vertical
\citep{Crouch+Cally2003, Cally2006, Schunker+Cally2006}. Similar
behavior is confirmed also in a 3D analysis by
\citet{Cally+Goossens2008}. The $p$-mode transformation at higher
frequencies is found to be significantly enhanced by moderate
inclinations  \citep{Crouch+Cally2003}. The explanation of this
effect is offered by the ray theory. The fast-mode high-$\beta$
waves (analog of $p$ modes) launched from their sub-photospheric
lower turning points reach the transformation layer with an angle
close to 20--30 degrees \citep{Schunker+Cally2006}. Thus, for the
moderately inclined magnetic field the attack angle $\psi$ in
Eq.~\ref{eq:cally} is small and the transformation is efficient.
Note that, \citet{Crouch+Cally2003, Schunker+Cally2006,
Cally+Goossens2008} discuss the case where the slow modes produced
after the transformation propagate upwards, while the downward
propagation is observed in simulations by
\citet{Cally+Bogdan+Zweibel1994, Cally+Bogdan1997,
Rosenthal+Julien2000}.

The critical role of the magnetic field inclination for the mode
transformation is confirmed by the numerical simulations of MHD
wave propagation in the chromosphere by
\citet{Carlsson+Bogdan2006}, where the authors considered
oscillations with wavelengths comparable to magnetic field scales.
At angles smaller than 30 degrees, much of the high-$\beta$
acoustic-like fast mode power is transformed into the low-$\beta$
slow modes propagating along the magnetic field lines. At larger
inclination angles the high-$\beta$ fast modes are refracted and
reflected and return back to the photosphere. Note that the larger
inclinations correspond to the regions where the field strength is
smaller and the transformation layer is located higher in the
atmosphere. When observed at particular  height, the interference
pattern produced by the upward and downward propagating waves in
these regions creates a ring of enhanced power around the magnetic
field concentrations \citep{Carlsson+Bogdan2006} which can be an
explanation of the acoustic halos observed at edges of active
regions (Sect. 4).

\section{Local helioseismology in active regions}

Time-distance helioseismology makes use of wave travel times
measured for wave packets traveling between various points on the
solar surface through the interior \citep{Duvall+etal1993,
Kosovichev1999, Kosovichev2002, Kosovichev+etal2000,
Zhao+Kosovichev2003}. The variations of these travel times
compared to the quiet unperturbed atmosphere are assumed to be
mainly due to mass flows and wave speeds below the surface. The
interpretation of the inversion results of time-distance
helioseismology encountered major critics when applied to magnetic
active regions of the Sun. Since magnetic field modifies the wave
propagation speeds in a similar way as the temperature
perturbations do, the magnetic and temperature effects on the wave
travel times are difficult to separate \citep[see
\eg][]{Moradi+Cally2008}. From observations, it has been
demonstrated repeatedly that waves travel 20--40 seconds faster
deep below the sunspots, when compared to quiet photosphere
measurements \citep[\eg][]{Duvall+etal1993, Braun1997}. The
explanation of this effect is far from clear. To understand the
influence of the magnetic field on travel time measurements, the
forward modeling of waves in magnetic regions has become the
preferred approach in recent years
\citep{Parchevsky+Kosovichev2008, Hanasoge2008, Cameron+etal2008,
Moradi+Cally2008, Moradi+etal2008, Khomenko+etal2008b}.

\citet{Parchevsky+Kosovichev2008} obtained that only 25\% of the
observed travel time difference can be explained by the magnetic
field effects in their simulations of waves excited by an isolated
source in an inclined magnetic field. Other authors report values
of the travel time differences from simulations that are similar
to the observed ones \citep{Cameron+etal2008, Moradi+Cally2008,
Moradi+etal2008, Khomenko+etal2008b}. The dependence of the
magnitude of the travel time difference on the magnetic field
strength was shown by \citet{Cameron+etal2008} and
\citet{Khomenko+etal2008b}, while \citet{Moradi+etal2008} argue
travel time differences are largely insensitive to the sunspot
structure. The frequency dependence of the travel time
measurements reported in the literature
\citep{Couvidat+Birch+Kosovichev2006, Couvidat+Rajaguru2007,
Rajaguru2008, Braun+Birch2008} is interpreted as one of the
demonstrations of the surface magnetic field effects. High
frequency waves have, in general, larger travel-time differences
\citep{Rajaguru2008}. Qualitatively, such dependence is reproduced
by numerical simulations as well as the ray theory eikonal
approach \citep{Moradi+etal2008, Khomenko+etal2008b}. The overall
agreement between the simulations and observations indicates that
the observed time-distance helioseismology signals in sunspot
regions correspond to fast high-$\beta$ MHD waves (modified $p$
and $f$ modes). This conclusion apparently contradicts the
chromospheric measurements, where the correlations between the
photospheric and chromospheric signals was found and the
longitudinal propagation of the low-$\beta$ slow waves allowed the
best agreement with the observations \citep{Centeno+etal2006a,
Bloomfield+etal2007}. In the same vein, \citet{Schunker+etal2005,
Schunker+Cally2006} argue that the observed dependence of the
Doppler signal on the line-of-sight magnetic field inclination
angle requires an alignment between the magnetic field direction
and the wave propagation, only possible for low-$\beta$ slow
waves. More work in needed to clarify these issues including the
calculation of the line formation heights and sunspot Wilson
depression effect.

\section{Conclusions}

Theoretical interpretation of waves observed in sunspots has made
big advances during the last decades. The influence of the
$\beta=1$ level and the magnetic field inclination on the observed
wave properties is being clarified, allowing the interpretation of
the observed oscillations in term of MHD wave modes. The observed
wave pattern in the upper photosphere and the chromosphere is
found to be compatible with the low-$\beta$ acoustic-like slow
mode propagation along the inclined magnetic field lines. Several
physical effects are proposed to be responsible for the frequency
change with height in the umbra from the 3 mHz in the photosphere
to 5--6 mHz in the chromosphere. The mode transformation at the
$\beta=1$ level is proposed to be responsible (at least in part)
for the wave power distribution over the active regions, including
the wave suppression in the umbra and, possibly, enhanced power of
acoustic halos. Theoretical interpretation of the local
helioseismology measurements including magnetic field has been
initiated. The first simulations of helioseismic waves in sunspots
show a general agreement of the wave travel times of the
high-$\beta$ fast modes with observations. Still, the final issue
in this field is far from clear and the major developments are
expected in the future. The 3-dimensional wave propagation and
mode transformation in sunspots are important to analyze from
simulations, including the chromospheric layers.

\acknowledgements %%% Text of acknowledgements runs on after this command.
This research has been funded by the Spanish Ministerio de
Educaci{\'o}n y Ciencia through project AYA2007-63881 and
AYA2007-66502.

%\aareferences

\end{document}